\documentclass[aps,nofootinbib,superscriptaddress,pre]{revtex4}
\usepackage{epsfig,graphics,amssymb,amsmath,subeqnarray,amsthm,latexsym}
\usepackage[colorlinks=true, citecolor=red, linkcolor=blue,urlcolor=blue ]{hyperref}
\usepackage{cleveref}
\usepackage{soul}
\usepackage[usenames,dvipsnames]{color}

\def\Re{{\rm Re}}
\def\Bo{{\rm Bo}}
\def\Ca{{\rm Ca}}

\def\e{{\bf e}}\def\q{{\bf q}}
\def\dd{\mathrm{d}}
\def\u{{\bf u}}\def\x{{\bf x}}
\def\M{{\bf M}}
\def\q{{\bf q}}
\def\f{{\bf f}}

\def\A{\mathcal{A}}
\def\R{\mathcal{R}}
\def\p{\partial}
\def\x{\mathbf{x}}
\def\n{\mathbf{n}}

\def\U{\mathbf{U}}
\def\m{{\rm m}}
\def\N{{\rm N}}
\def\Pa{{\rm Pa}}
\def\mol{{\rm mol}}
\def\kg{{\rm kg}}
\def\K{{\rm K}}
\def\J{{\rm J}}
\def\s{\rm{s}}

\def\Pe{{\rm Pe}}

\begin{document}
\title{Physics of Bubble-Propelled Microrockets}
\author{Giacomo Gallino}
\email{giacomo.gallino@epfl.ch}
\affiliation{Laboratory of Fluid Mechanics and Instabilities, EPFL, CH-1015 Lausanne, Switzerland.}
\affiliation{Department of Applied Mathematics and Theoretical Physics, University of Cambridge, CB3 0WA, United Kingdom}
\author{Fran\c cois Gallaire}
\email{francois.gallaire@epfl.ch}
\affiliation{Laboratory of Fluid Mechanics and Instabilities, EPFL, CH-1015 Lausanne, Switzerland.}
\author{Eric Lauga}
\email{e.lauga@damtp.cam.ac.uk}
\affiliation{Department of Applied Mathematics and Theoretical Physics, University of Cambridge, CB3 0WA, United Kingdom}
\author{Sebastien Michelin}
\email{sebastien.michelin@ladhyx.polytechnique.fr}

\affiliation{LadHyX -- D\'epartement de M\'ecanique,\\ CNRS -- Ecole Polytechnique, 91128 Palaiseau Cedex, France}
\affiliation{Department of Applied Mathematics and Theoretical Physics, University of Cambridge, CB3 0WA, United Kingdom}
\date{\today}
\begin{abstract}

A popular method to induce synthetic propulsion at the microscale is to use the forces created by surface-produced gas bubbles inside the asymmetric body of a catalytic swimmer (referred to in the literature as microrocket). Gas bubbles nucleate and grow within the catalytic swimmer and migrate toward one of its opening under the effect of asymmetric geometric confinement, thus generating a net hydrodynamic force which propels the device. In this paper we use numerical simulations to develop a joint chemical (diffusive) and hydrodynamic (Stokes) analysis of the bubble growth within a conical catalytic microrocket and of the associated bubble and microrocket motion. Our computational model allows us to solve for the bubble dynamics over one full bubble cycle ranging from its nucleation to its exiting the conical rocket and therefore to identify the propulsion characteristics as function of all design parameters, including geometry and chemical activity of the motor, surface tension phenomena, and all physicochemical constants. Our results suggest that hydrodynamics and chemistry partially decouple in the motion of the bubbles, with hydrodynamics determining the distance travelled by the microrocket over each cycle while chemistry setting the bubble ejection frequency. Our numerical model finally allows us to identify an optimal microrocket shape and size for which the swimming velocity (distance travelled per cycle duration) is maximized.

\end{abstract} 

\maketitle


\section{Introduction}
Artificial microswimmers have recently attracted much attention across many scientific disciplines: from a fundamental point of view, they represent alternatives to biological systems (e.g.~bacteria, algae) to characterize and control individual propulsion at the micron scale and collective organization in so-called active fluids. They also present many opportunities for engineering applications, in particular in the biomedical context to perform such tasks as drug delivery~\cite{gao2014synthetic,wang2012nano}, nanosurgery~\cite{nelson2010microrobots}, cell sorting~\cite{solovev2010magnetic,garcia2013micromotor}. Two of the main challenges are to overcome the restrictions inherent to propulsion in highly-viscous environments such as reversibility and symmetry-breaking~\cite{lauga2009hydrodynamics} as well as miniaturization.

Proposed designs can currently be classified into two broad categories, namely (i) \emph{actuated} systems which rely on an externally-imposed forcing, most often at the macroscopic level, in order to self-propel (e.g.~an unsteady magnetic or acoustic field~\cite{dreyfus2005,ghosh2009,wang2012}) and (ii) \emph{catalytic} (or fuel-based) systems which rely on local physico-chemical processes (e.g.~chemical reactions at their surface) to convert chemical energy into a mechanical displacement~\cite{ebbens2016,duan2015,yadav2015,xu2017light}. For the latter, this energy conversion may follow different routes, a popular one being the generation of gas bubbles whose growth and dynamics enable propulsion~\cite{li2016rocket}.

These so-called microrockets represent one of the most promising designs for applications. In contrast with active phoretic colloids~\cite{paxton2004,howse2007,volpe2011,moran2017phoretic} that swim exploiting local physico-chemical gradients in order to generate hydrodynamic forcing~\cite{anderson1989colloid}, microrockets move due to the production of gas bubbles inside the asymmetric body of the swimmer. More precisely, bubbles nucleate and grow within the catalytic motor and migrate toward one of its opening under the effect of the asymmetric geometric confinement, thus generating a net hydrodynamic force which propels the device.

Such configuration has been studied experimentally, focusing primarily on the rocket velocity and resulting trajectory~\cite{gao2012catalytically,mou2015single,zha2018tubular}. In parallel, the first in-vivo application of this technology for drug delivery was recently conducted~\cite{gao2015artificial}. Fundamental understanding is still needed of the role played by the different hydrodynamic and chemical mechanisms involved as well as their couplings, in order to identify optimal design rules for such microrockets. Several studies have proposed partial modeling of the problem, focusing more specifically on the motion of the bubble inside the rocket~\cite{manjare2013bubble,fomin2014propulsion} or during and after its ejection~\cite{li2011dynamics,li2014hydrodynamics}.

The purpose of the present work is to propose a detailed chemical and hydrodynamic analysis of the bubble growth within the catalytic microrocket, and associated bubble and microrocket motion, in order to identify the role of the different design parameters in setting the propulsion speed. To this end, we propose an accurate numerical simulation of the dissolved gas diffusion and fluid motion both inside and outside a conical catalytic microrocket, assuming that the predominance of capillary effects ensures that the bubble remains   spherical while translating inside the conical body. Focusing specifically on the bubble growth within the motor, we monitor the bubble dynamics over one bubble cycle ranging from its nucleation to its exiting the conical rocket. The propulsion characteristics are clearly identified in terms of the design characteristics (i.e.~geometry and chemical activity of the motor, surface tension phenomena, ambient conditions).

The bubble dynamics is observed to be composed of two different phases, which are characterized in detail in terms of bubble and microrocket motion as well as hydrodynamic signature, an information of particular importance which conditions the hydrodynamic interaction of the artificial swimmer with its environment (e.g.~boundaries or other swimmers) and critically influences its trajectory and its control. Our results further suggest that for such bubbles, hydrodynamics and chemistry partially decouple, the former determining the distance travelled by the microrocket over each cycle, while the latter determines the cycle duration or bubble ejection frequency. Moreover, we are able to identify an optimal microrocket shape and size for which the swimming velocity is maximized.

\section{Model}
\label{sec:models}

\begin{figure}

\center
\includegraphics[width=8cm]{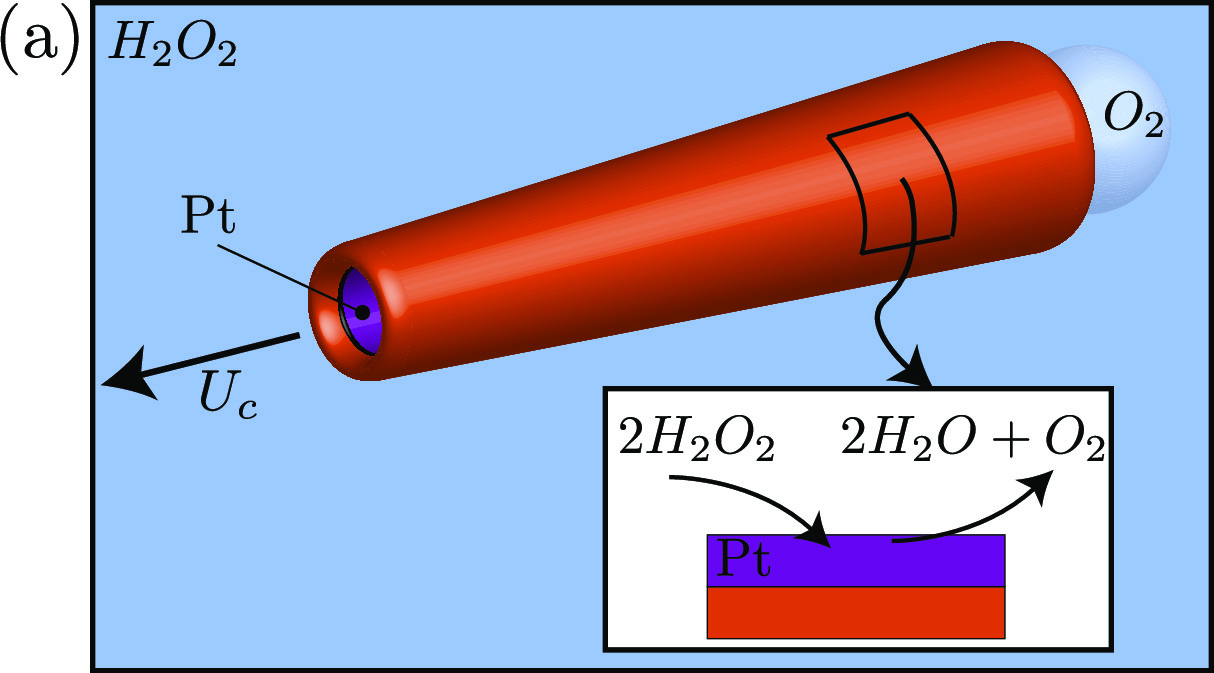}
\includegraphics[width=8cm]{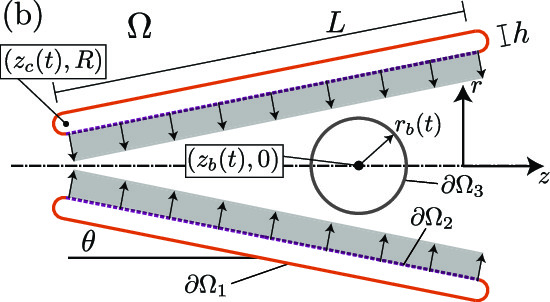}

\caption{Microrocket: (a) Dissolved oxygen ($O_2$) is produced by the decomposition of hydrogen peroxide onto a platinum-coated surface. The saturated oxygen environment in the confined motor leads to bubble nucleation and growth. Bubbles exit from the larger opening while the microrocket is propelled in the opposite direction. (b) Schematic of the problem and side view of the axisymmetric microrocket in a plane containing the central axis. Red solid surfaces are inert (no oxygen production) while reactive surface where oxygen emission occurs are shown as dashed purple lines.}
\label{fig:geometry}

\end{figure}

The dynamics of an isolated microrocket is considered here in a fluid of density $\rho = 10^{3}\, \kg \,\m^{-3}$ and dynamic viscosity $\mu = 10^{-3}\, \Pa \,\,\s$ (i.e.~water). The axisymmetric microrocket geometry is that of a cone of radius $R=1\,\mu\m$, aspect ratio $\xi=L/R$, thickness $h/R$ and opening angle $\theta$ (see Figure~\ref{fig:geometry}). Throughout this study we use cylindrical coordinates $(r,z)$ measured in units of cone radius $R$, and set $\xi=10$ and $h/R=0.2$, which are compatible with experimental designs~\cite{manjare2013bubble,li2011dynamics,li2016rocket}. The inner surface of the cone is chemically-active and catalyses a chemical reaction. One of the products of this reaction is a soluble gas. We pick $O_2$ as a specific example, produced by the decomposition of hydrogen peroxide on platinum, which diffuses in the liquid with molecular diffusivity $D = 2 \times 10^{-9} \,\m^2 \,\,\s^{-2}$; note that our modeling approach is generic and easily applicable to other chemical situations. The production of oxygen on the surface of the catalyst is modelled here as a fixed molar flux $\A = 10^{-2} \mol \,\, \m^{-2} \,\, \s$ (considering the experimental data in~\cite{manjare2013bubble}). The rate of production of the gas is large enough that the fluid is saturated so that a spherical gas bubble of radius $r_b(t)$ grows within the rocket. Throughout this study, we consider the dynamics of a bubble on the axis of symmetry (see Section~\ref{sec:bubble_nucleation} for more details). The viscosity and density of the gas  are negligible compared to that of the liquid, and, although in experimental conditions the surface tension value is often affected by the presence of surfactants, we initially consider $\gamma = 7.2 \times 10^{-2} \,\N\,\, \m^{-1}$. In section~\ref{sec:optimal}, we will then systematically vary the value of $\gamma$ in order to study the effect of surfactants, which play a crucial role in real applications by stabilizing the bubble.

Based on the above characteristics and the experimentally-measured microrocket velocity $U_c\sim 5\times 10^{-4}\m\,\,\s^{-1}$~\cite{manjare2013bubble,li2011dynamics,li2016rocket}, the effect of inertia and gravity can be neglected (the Reynolds $\Re=\rho U_c R/\mu\sim 5\times 10^{-4}$ and Bond numbers $\Bo=gR^2\rho/\gamma\sim 10^{-6}$ are both small). The Peclet number $\Pe=RU_c/D\sim 0.2$ is less than one and we may neglect gas advection and unsteady diffusion within the rocket as a first approximation. Finally, because the typical hydrodynamic stresses are negligible compared to those due to surface tension (the capillary number $\Ca=\mu U_c /\gamma\sim10^{-5}$ is small), the bubble is expected to remain spherical during its entire evolution and hydrodynamic effects do not contribute to the bubble inner pressure.

\subsection{Gas diffusion}

Following these dimensional considerations, the non-dimensional dissolved gas concentration $c(\x)$, relative to its far-field concentration and measured in units of $\A R/D$, satisfies the steady diffusion equation in the liquid domain $\Omega$,
\begin{equation}
\nabla^2 c = 0, \label{eq:laplace}
\end{equation}
and decays in the far-field ($c\rightarrow 0$ for $\x\rightarrow \infty$). The dissolved gas is produced through chemical reaction on the inner surface of the cone, and $c(\x)$ therefore also satisfies
\begin{align}
- \n \cdot \nabla c &= 0 \quad \text{for} \quad \x \in \p \Omega_1, \label{eq:bcFlux1} \\
- \n \cdot \nabla c &= 1 \quad \text{for} \quad \x \in \p \Omega_2. \label{eq:bcFlux2}
\end{align}
where $\p \Omega_1$ and $\p \Omega_2$ refer to the inert and active cone surfaces respectively (see Figure~\ref{fig:geometry}b). At the bubble surface, the dissolved gas is in thermodynamic equilibrium with the gas pressure inside the bubble, and $c$ is therefore given by Henry's law $c= H^{cc} p_b$, where $H^{cc}= H^{cp}D\gamma/R^2 \A$ is the non-dimensional volatility constant and $p_b$ is the pressure inside the bubble measured in units of $\gamma/R$. The volatility constant is an intrinsic property of the dissolved gas (e.g.~$H^{cp}=4\times 10^{-7}\kg\,\,\m^{-3}\,\,\mbox{Pa}^{-1}$ for oxygen in water~\cite{lv2017growth}). The bubble pressure $p_b$ is given by Laplace law ($\Ca\ll 1$), so that in non-dimensional form
\begin{equation}
c = {H^{cc}} \left( \beta +\frac{2}{r_b} \right) \quad \text{for} \quad \x \in \p \Omega_3, \label{eq:henry}
\end{equation}
where $\beta=\tilde{p}_0 R/\gamma$ is the ratio between the dimensional ambient pressure $\tilde{p}_0$ and capillary pressure. Equations~\eqref{eq:laplace}--\eqref{eq:henry} determine $c(\x)$ uniquely. Using this solution, the flux of dissolved gas into the bubble is computed as
\begin{equation}
Q = \int_{\p \Omega_3} (\n \cdot \nabla c) \,\dd S.
\end{equation}
Using mass conservation of the gas species and the ideal gas law, the time evolution of the bubble radius is finally determined in non-dimensional form as
\begin{align}
\frac{dr_b}{dt} &= \frac{Q}{3\beta r_b^2+4 r_b}, \label{eq:growth}
\end{align}
where the time is measured in units of $4\pi \gamma/3\A \R T_0$, $\mathcal{R}=8.314 \,\, \J \,\, \mol^{-1} \K^{-1}$ is the universal gas constant and $T_0 \sim 293 \,\, \K$ is the ambient (room) temperature.

\subsection{Hydrodynamics}

The change in bubble size imposed by the chemical reaction and gas diffusion dynamics, equation~\eqref{eq:growth}, sets the fluid into motion within the rocket. Since $\Re\ll 1$, the non-dimensional fluid velocity $\u$ and hydrodynamic pressure field $p$ (now measured in units of $3\A \R T_0 R/4\pi \gamma$ and $3 \mu \A \R T_0/4\pi \gamma$, respectively) satisfy the incompressible Stokes' equations in the liquid domain $\Omega$
\begin{align}
-\nabla p + \nabla^2 \u = 0, \quad \nabla \cdot \u = 0.
\end{align}
The cone and bubble are translating along the axis of symmetry with respective velocities $\U_c = \dot z_c\e_z$ and $\U_b=\dot z_b\e_z$, with $z_c(t)$ and $z_b(t)$ the axial positions of the cone's narrow opening and of the bubble center, respectively. One of the main purposes of this paper is to compute the time-dependent values of both $\U_c$ and $\U_b$. At the surface of the solid cone, the no-slip boundary condition imposes the boundary condition
\begin{equation}
\u = \U_c \quad \text{for} \quad \x \in \p \Omega_1,\p \Omega_2, \label{eq:bcCone}
\end{equation}
and the hydrodynamic flow decays in the far-field ($\u\rightarrow 0$ and $p\rightarrow p_0$ as $\x \rightarrow \infty$). The bubble is inflating while translating, and its surface is free of any tangential stress, so that mixed boundary conditions must be satisfied 
\begin{equation}
\u \cdot \n = \U_b \cdot \n +\frac{dr_b}{dt}\quad \textrm{and } \quad (\mathbf{I}-\mathbf{n}\mathbf{n}) \cdot \boldsymbol \sigma \cdot \n =0 \quad \text{for} \quad \x \in \p \Omega_3, \label{eq:bcBubble}
\end{equation}
with $\mathbf{n}$ the normal unit vector to the bubble surface. The coupling between chemistry and hydrodynamics enters only in equation~\eqref{eq:bcBubble} through the inflation rate. The problem is closed by imposing that the cone and the bubble are each force-free during their axisymmetric translation along the axis (inertia is negligible here, as explained previously)
\begin{align}
\int_{\p \Omega_1+\p \Omega_2} (\n \cdot \boldsymbol \sigma \cdot \e_z ) \,\dd S =\int_{\p \Omega_3} (\n \cdot \boldsymbol \sigma \cdot \e_z ) \,\dd S = 0,\label{eq:forcefree}
\end{align}
a closure relationship which implicitly determines the instantaneous values of the bubble and cone velocities.

\subsection{Numerical method and validation}

Both the diffusion (Laplace) and hydrodynamic (Stokes) equations are solved numerically using axisymmetric Boundary Element Methods. The boundary integral equation for the axisymmetric solute concentration on the boundaries is classically written for $\x_0\in\p\Omega$ as (see equation $(4.5.5)$ in Ref.~\cite{pozrikidis2002practical})
\begin{equation}
c(\x_0) = -2\int_{\p \Omega} G(\x,\x_0) [\n(\x) \cdot \nabla c(\x)] r(\x) \dd l (\x) + 2\int_{\p \Omega}^{PV} c(\x) [\n(\x) \cdot \nabla G(\x,\x_0)] r(\x) \dd l (\x), \label{eq:laplaceBIE}
\end{equation}
where $G(\x,\x_0)$ is the axisymmetric Green's function of the Laplace equation (see equation (4.5.6) in Ref.~\cite{pozrikidis2002practical}), $PV$ denotes the principal-value integral and $\n$ is the normal vector pointing into the liquid domain $\Omega$, whose boundaries $\p\Omega$ consists in the cone and bubble surfaces. Discretizing these boundaries into $N$ piecewise constants elements, and applying boundary conditions, equations~\eqref{eq:bcFlux1} and~\eqref{eq:bcFlux2}, equation~\eqref{eq:laplaceBIE} provides a $N\times N$ linear system for the value of the concentration on each element, which is solved using classical matrix inversion techniques.

Similarly, using the fundamental integral representation of Stokes flows~\cite{pozrikidis1992boundary}, the fluid velocity $\u$ is expressed on the boundaries $\p\Omega$ as
\begin{align}
4 \pi \u(\x_0) = -\int_{\p \Omega} \M(\x,\x_0) \cdot \f(\x) dl(\x) + \int_{\p \Omega}^{PV} \n(\x) \cdot \q(\x,\x_0) \cdot \u(\x) dl(\x), \label{eq:stokesBIE}
\end{align}
 where $\M$ and $\q$ are the axisymmetric Stokeslet and associated stress respectively (we follow the notation in~Ref.~\cite{pozrikidis1992boundary}) and $\f=\boldsymbol \sigma \cdot \n$ is the traction acting on the boundaries. Equation~\eqref{eq:stokesBIE} can be rewritten more conveniently on the cone and bubble surface respectively, by using the no-slip and mixed boundary conditions respectively, equations~\eqref{eq:bcCone} and \eqref{eq:bcBubble}, as~\cite{pozrikidis2005computation,zhu2013low}
\begin{align}
4 \pi \u_\text{rel} (\x_0) &= -\int_{\p \Omega} \M(\x,\x_0) \cdot \f(\x) dl(\x) + \int_{\p \Omega_3}^{PV} \n(\x) \cdot \q(\x,\x_0) \cdot \u_\text{rel} (\x) dl(\x) - 8 \pi \U_b \label{eq:stokesBIE1}\\
&\qquad \text{for} \quad \x_0 \in \p \Omega_3, \notag \\
0 &= -\int_{\p \Omega} \M(\x,\x_0) \cdot \f(\x) dl(\x) + \int_{\p \Omega_3} \n(\x) \cdot \q(\x,\x_0) \cdot \u_\text{rel} (\x) dl(\x) - 8 \pi \U_c \label{eq:stokesBIE2}\\
&\qquad \text{for} \quad \x_0 \in \p \Omega_1+\p \Omega_2,\notag
\end{align}
where $\u_\textrm{rel}$ is the relative velocity of the bubble surface to its center of mass whose normal component is set by the bubble growth rate, $\u_\textrm{rel}\cdot\n=\dot{r}_b$. The principal value integral appears only in equation~\eqref{eq:stokesBIE1}, when $\x_0$ belongs to the integration path $\p \Omega_3$.

When discretizing the boundaries into $N$ piecewise constant elements, equations~\eqref{eq:stokesBIE1} and \eqref{eq:stokesBIE2}, together with the force-free conditions equation~\eqref{eq:forcefree}, provide a $(2N+2)\times (2N+2)$ for (i) the axial and radial components of the fluid traction on the cone surface, (ii) the tangential relative fluid velocity and normal traction on the bubble surface and (iii) the bubble and cone axial velocities. A particular technical point deserves special attention: when $\x \rightarrow \x_0$ the Green's function in equations~\eqref{eq:laplaceBIE},~\eqref{eq:stokesBIE},~\eqref{eq:stokesBIE1} and~\eqref{eq:stokesBIE2} become singular and special (but classical) treatment is needed in order to maintain good accuracy~\cite{pozrikidis2002practical,pozrikidis1992boundary}.

The Laplace and Stokes solvers described above were validated by computing the chemical field around a Janus particle and its swimming velocity due to diffusiophoretic effects and a good agreement was found with the analytical solution~\cite{michelin2014phoretic}.

The bubble is growing within a confined environment. As observed in Section~\ref{sec:results}, the liquid gap between the bubble and cone may become small during the bubble formation and expulsion. Adaptive mesh refinement is therefore needed to maintain sufficient numerical accuracy: the elements are split into two when their size is larger than the gap. However, accurately resolving the hydrodynamic stresses when the bubble approaches the cone wall would require a prohibitive number of mesh elements for thin gaps. The physical effect of these hydrodynamic lubrication stresses is however essential to prevent overlap between the bubble and cone surfaces. We therefore introduce short-ranged repulsive forces to prevent such overlap numerically. The force-free conditions along the axial direction now write
\begin{equation}
F=2\pi \int_{\p \Omega_1+\p \Omega_2} r (\n \cdot \boldsymbol \sigma \cdot \e_z)\,\dd l = -2\pi \int_{\p \Omega_3} r (\n \cdot \boldsymbol \sigma \cdot \e_z)\,\dd l,
\end{equation}
where, similarly to what was done in Ref.~\cite{freund2007leukocyte}, the axial repulsive force $F$ is defined as
\begin{equation}
F=
\begin{cases}
\displaystyle{B\frac{e^{\delta-d}-1}{e^{\delta}-1} \frac{|\mathbf{d} \cdot \e_z|}{d}} \quad & \text{for} \quad d \leq \delta \\ \displaystyle
0 \quad & \text{for} \quad d > \delta
\end{cases}
\end{equation}
where $\mathbf{d}$ is the minimum distance vector between the bubble and the cone and $d=||\mathbf{d}||$. This simply adds a repulsive interaction between the bubble and cone and the global system remains overall force-free. In the following, $B=10^{7}$ and $\delta=0.1$ are used; varying these parameters does not significantly affect the numerical results. The system of first order differential equations for $r_b(t)$, $z_c(t)$ and $z_b(t)$ is marched in time using Matlab's ODE23t routine, which uses a semi-implicit, adaptive time-stepping scheme.

\section{Results}
\label{sec:results}
\subsection{Dissolved gas distribution and bubble growth}
\label{sec:bubble_nucleation}

\begin{figure}[h]
\center
\includegraphics[width=\textwidth]{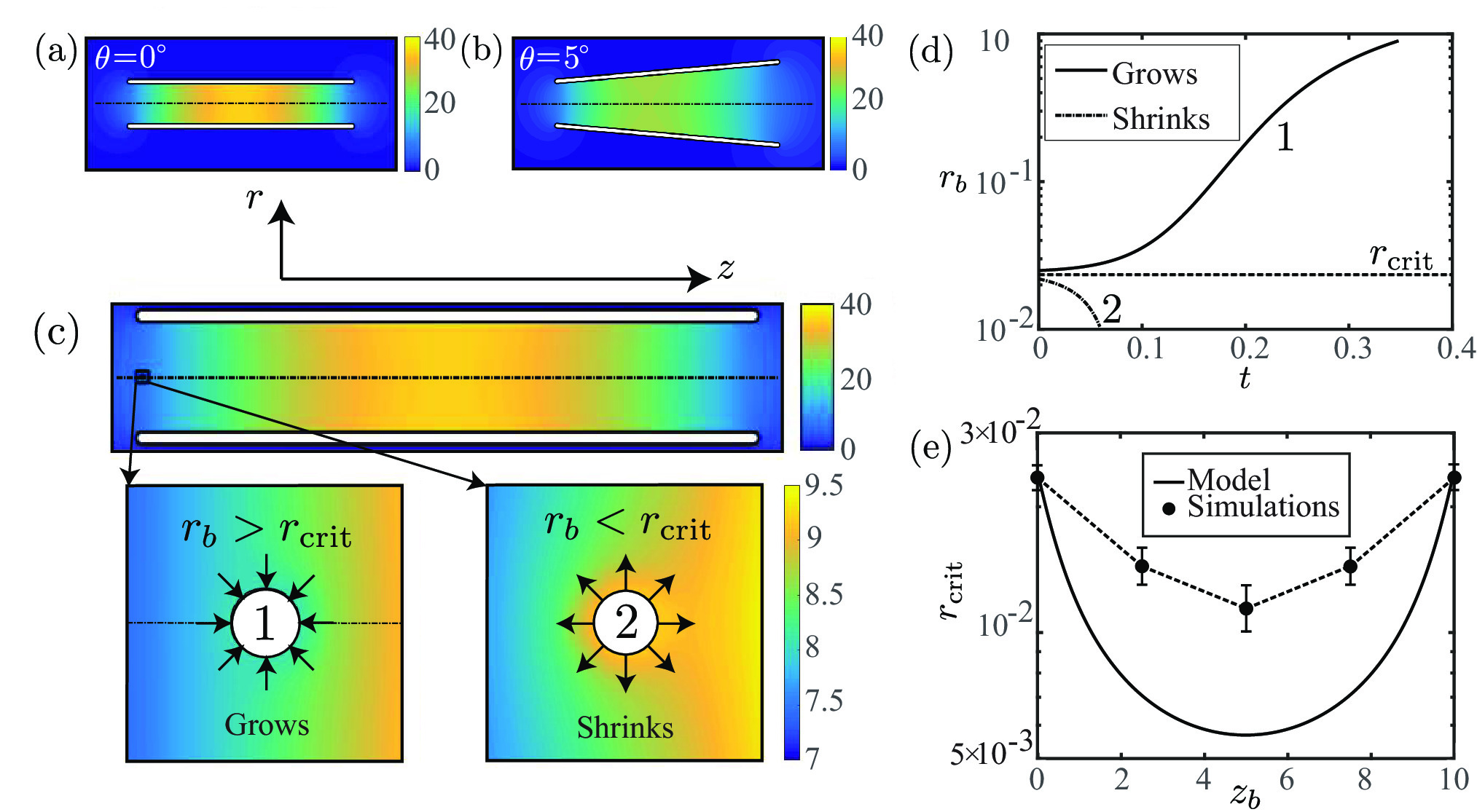}
\caption{Dissolved gas concentration inside and around the microrocket for (a) $\theta=0^\circ$ and (b) $\theta=5^\circ$. (c) Effect of the presence of the bubble on the local dissolved gas concentration for $\theta=0^\circ$. The top panel is the concentration with no bubble while the central and bottom panels show the concentration around a growing (1) or shrinking (2) bubble, and (d) the associated evolution of the bubble radius. The black arrows indicate the direction of the diffusive flux of dissolved gas. (e) Critical radius as a function of bubble position along the axis for $\theta=0^\circ$ obtained numerically results (with error bars) or from the estimation $r_\text{crit}=2H^{cc}/(c(z,0)-\beta H^{cc})$. For all panels, $H^{cc}=0.1$ and $\beta=1$.}
\label{fig:chemicalField}
\end{figure}

When $\theta=0^\circ$, the microrocket is cylindrical and the concentration of dissolved gas is left-right symmetric with a maximum in the center of the microrocket due to the geometric confinement (Figure~\ref{fig:chemicalField}a). When $\theta\neq 0^\circ$, the left-right symmetry is broken and the maximum concentration moves toward the smaller cone opening (Figure~\ref{fig:chemicalField}b). It is worth noting that the overall concentration level becomes smaller due to the increased dissolved gas diffusion out of the cone resulting from the weaker confinement.

We are now interested in understanding how the chemical field generated by the microrocket impacts and controls the growth of a bubble. To this end, it should be reminded that a gas bubble placed in a uniform concentration of dissolved gas shrinks (resp.~grows) when its surface concentration $c_b$ is higher (resp.~lower) than the background concentration $c_\infty$. Namely, when $c_b<c_\infty$ the diffusive flux of gas species is oriented toward the bubble and vice versa. Since $c_b$ decreases as the bubble radius $r_b$ increases, see equation~\eqref{eq:henry}, large bubbles are more likely to grow than small bubbles.

In the microrocket geometry, a bubble located outside the cone will shrink more easily than one located inside, due to the lower level of gas concentration outside the rocket. But even when a bubble is inside the conical microrocket, it is still expected to shrink if its concentration is higher than the background concentration established by the microrocket (e.g. for small enough bubbles). The concentration of the bubble surface is set by the thermodynamic equilibrium at the bubble surface (Henry's law, equation~\ref{eq:henry}) and it increases when $H^{cc}$ (inversely proportional to the surface flux $\A$) or $\beta$ increase. Based on experimental estimates~\cite{manjare2013bubble,li2014hydrodynamics}, we set $H^{cc}=0.1$ and $\beta=1$ and numerically compute the net flux of dissolved gas into the bubble in order to determine whether a bubble will grow or shrink depending on its radius and axial position. For each bubble position, a critical radius $r_\textrm{crit}(z_b)$ is identified as the minimum radius for which bubble growth is observed at a fixed location (Figure~\ref{fig:chemicalField}e).

Throughout this study, an axisymmetric problem is considered where the bubble center is located on the axis of symmetry. This assumption seems reasonable when considering an inertialess bubble translating in a channel \cite{rivero2017bubbles}, although it neglects the initial bubble migration from the catalyst surface, where it most likely nucleates, toward the axis. Bubble nucleation on a catalyst surface considered in some recent studies~\cite{lv2017growth} is a complex physico-chemical phenomenon which is beyond the scope of the present study, that focuses on the coupling of gas diffusion and hydrodynamics resulting in the rocket propulsion.

Figure~\ref{fig:chemicalField}c shows the dissolved gas concentration for a bubble of radius slightly larger (resp.~smaller) than $r_\textrm{crit}$, corresponding to a growing (resp.~shrinking) bubble. The gas concentration around the bubble is locally lower (resp.~higher) due to the bubble presence, as a result of the diffusive flux of gas toward (resp.~away from) the bubble (the flux direction is indicated with arrows in the lower panels of Figure~\ref{fig:chemicalField}c).

The critical radii found numerically are small compared to the cone size. Assuming further that the bubble is small compared to the local length scale for the gas concentration changes with no bubble, the bubble is expected to grow if its surface concentration, equation~\eqref{eq:henry}, is lower than the local background concentration (i.e.~when the bubble is not present). This provides an estimate of the critical radius as $r_\text{crit}=2H^{cc}/(c(z,0)-\beta H^{cc})$. This estimate agrees qualitatively with the numerical solution (see Figure~\ref{fig:chemicalField}e): the estimated critical radius is of the same order as the numerical solution, and is smaller in the middle of the motor, where the concentration is higher. The quantitative discrepancy most likely arises from finite-size effects of the bubble: while critical radii are small compared to the cone radius, they are not negligible over the characteristic length scale introduced by the local gas concentration gradients within the conical motor. 

In the following simulations, the initial bubble conditions (radius and position) are chosen as those for the smallest bubble that can grow: its position $z_b(0)$ and radius $r_b(0)$ are identified by the location of the minimum of the critical radius $r_\textrm{crit}$ along the axis and the corresponding critical radius. A small constant $C_0=0.1$ is added to $r_b(0)$ in order to ensure bubble growth and avoid spurious bubble shrinking due to numerical inaccuracy.

\subsection{Definition and dynamics of the bubble cycle}
\label{sec:bubble_cycle}
\begin{figure}[h]
\center
\includegraphics[width=\textwidth]{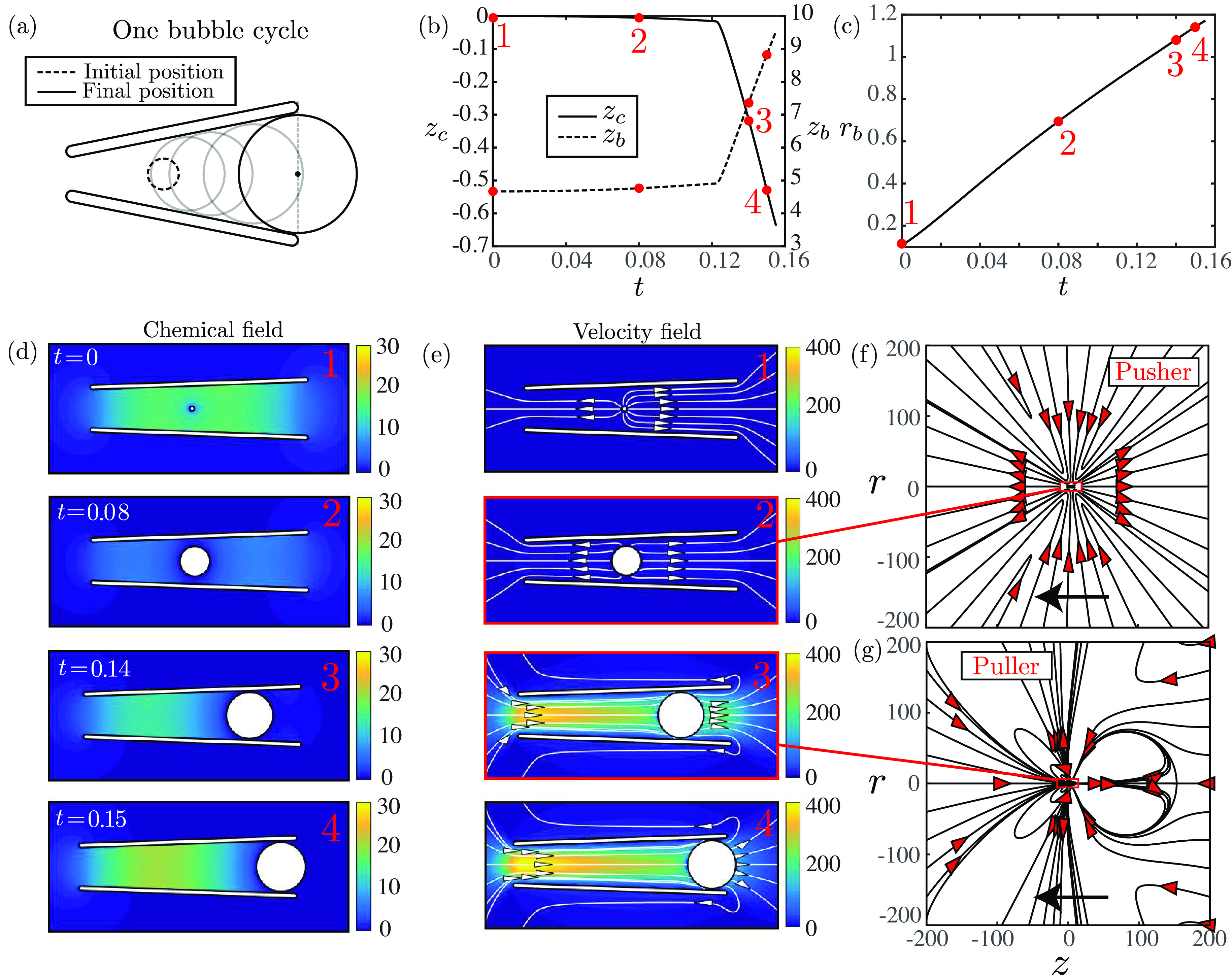}
\caption{Chemical and bubble dynamics over one bubble cycle for $H^{cc}=0.1$, $\beta=1$ and $\theta=2^\circ$: (a) Sketch of one bubble cycle, starting when the bubble is in the initial position and ending when it exists the cone. (b,c) Time-dependence of the cone and bubble positions. (d) Time-dependence of the bubble radius $r_b$. (d,e) Snapshots of the dissolved gas concentration and velocity field (streamlines and intensity) for the four instants in panels (b,c). (f,g) Large scale velocity field which is a pusher/puller when the bubble is non-confined/confined while the arrows indicate the swimming direction.}
\label{fig:snapshots}
\end{figure}

Starting from the initial conditions described in the previous section, the diffusion and hydrodynamic equations are solved numerically, and the bubble and motor displacements are investigated during a single \emph{bubble cycle}, i.e.~the growth of a single bubble. This bubble cycle is defined starting from the initial condition described previously (nucleation) and finishing when the bubble center exits the cone (Figure~\ref{fig:snapshots}a). The bubble cycle is shown in video $\#1$ of the Supporting Information.

The evolution of the bubble and cone displacements $z_b(t)$ and $z_c(t)$, as well as that of the bubble radius $r_b(t)$, are shown over one bubble cycle in Figure~\ref{fig:snapshots}b-c for fixed $H^{cc}$, $\beta$ and cone geometry.
The bubble is initially small and not confined by the cone geometry. It is growing, thanks to the absorption of dissolved gas by diffusion at its surface, thereby lowering the concentration in its vicinity. In this first, not geometrically-confined phase, the bubble growth pushes fluid out through the small and large openings of the cone and both the cone and bubble displacements are small. Hydrodynamically, the microroket is a so-called ``pusher'' during this phase (see Figure~\ref{fig:snapshots}f). Like swimming bacteria, it pushes fluid away along its axis of symmetry while pumping fluid toward it in the equatorial plane~\cite{lauga2009hydrodynamics}. 

A transition to a second phase is observed for $t\approx 0.12$ when the bubble has grown sufficiently for the confinement by the walls of the motor to become significant. As it continues growing under the effect of the dissolved gas diffusion, the bubble translates rapidly toward the larger opening. Figure~\ref{fig:snapshots}b shows that most of the cone displacement occurs during this second phase. Because of the fast relative translation of the confined bubble with respect to the motor, fluid is sucked in from the smaller opening and pushed out of the rocket at the larger opening (Figure~\ref{fig:snapshots}e). Overall, the hydrodynamic signature of the microrocket is reversed in this phase as it now acts as a so-called ``puller'' although 
higher-order contributions to the hydrodynamic signature create more complex flow structures in the vicinity of the rocket such as the recirculation zone in the back (see Figure~\ref{fig:snapshots}g). We conjecture that these different (and unsteady) flow signatures might play an important role in the flow-mixing generated by the displacement of the microrocket ~\cite{orozco2014bubble}. When the bubble exits the cone, the bubble cycle ends. Following the bubble motion toward the exit, the concentration of dissolved gas within the microrocket recovers its initial levels, allowing for a new bubble cycle (Figure~\ref{fig:snapshots}d).

\subsection{Influence of physico-chemical properties on the microrocket displacement}
\label{sec:results_chem}
\begin{figure}[h]
\center
\includegraphics[width=\textwidth]{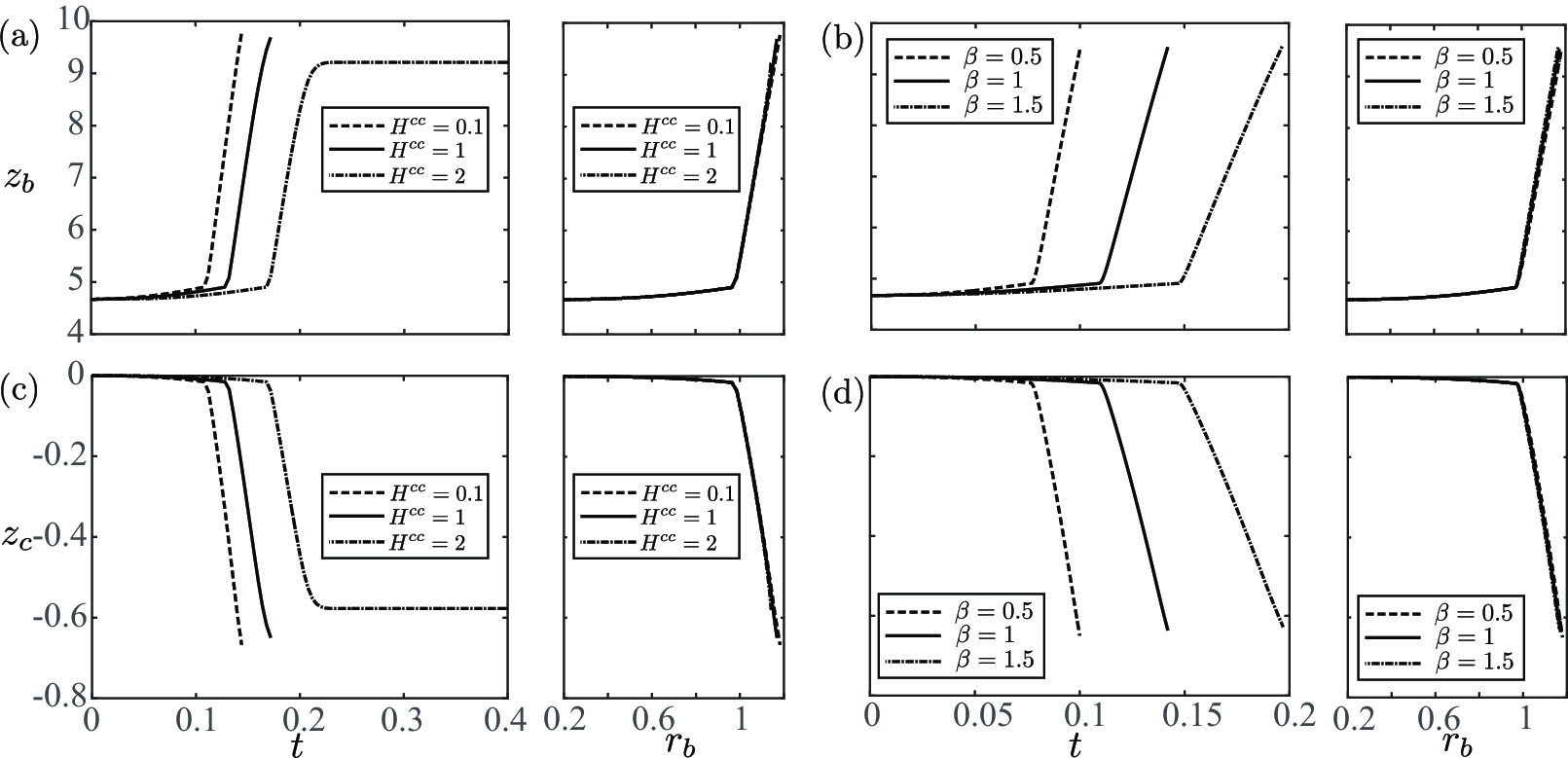}
\caption{Influence of the physico-chemical parameters for $\theta=2^\circ$. Evolution in time of (a,b) the bubble position $z_b(t)$ and (c,d) the cone position $z_c(t)$ for different values of (a,c) $H^{cc}$ and (b,d) $\beta$. In (a,c) $\beta=1$ and in (b,d) $H^{cc}=0.1$. In each panel, the figure on the right show the evolution of $z_b$ or $z_c$ with respect to $r_b$.}
\label{fig:parametricChemicals}
\end{figure}

As emphasized above, the chemical properties of the catalyst and gas species (e.g.~the flux of dissolved gas $\A$ or it volatility $H^{cp}$) and the background pressure $\tilde{p}_0$ influence the magnitude of the dissolved gas concentration within the microrocket and at the bubble surface. The effect of such quantities, and of their non-dimensional counterpart $H^{cc}$ and $\beta$, on the motor and bubble dynamics is investigated in Figure~\ref{fig:parametricChemicals} for a given rocket geometry.

Increasing the value of $H^{cc}$ (see Figure~\ref{fig:parametricChemicals}a and \ref{fig:parametricChemicals}c) is equivalent to reducing the chemical activity of the catalyst, $\A$, or increasing the gas volatility. For instance, recent experimental studies have shown that enhanced surface activity can be achieved by varying the roughness of the catalysts~\cite{li2014hierarchical,maria2016carbon}. Although we consider in our model only smooth surfaces, such effects could be described, as a first approximation, by a decrease of the effective value of $H^{cc}$. Figure~\ref{fig:parametricChemicals}a and c show that the total displacement of the bubble over one cycle is almost unchanged when $H^{cc}$ is varied. Changing $H^{cc}$ modifies however the duration of the bubble cycle. Moreover, for $H^{cc}=2$, the bubble stops inflating before leaving the cone (the diffusive flux of gas at its boundary is not sufficiently large); in that case, the bubble cycle is not closed and the microrocket will not be able to reach a continuous motion. Similarly, the time required for a full bubble cycle is observed to increase with $\beta$, but this does not significantly affect the total displacement (Figure~\ref{fig:parametricChemicals}b,d).
 
This independence of the kinematic displacement from the chemical characteristics is an illustration of the decoupling between the chemical and hydrodynamic problems due to the negligible deformation of the spherical bubble. The shape of the bubble is here solely described by the growth of its radius, and because the bubble grows monotonously, a bubble cycle can be parameterized by the bubble size (rather than time) starting from the initial critical radius and up to its final radius at the exit. At leading order, the latter is solely determined by the cone geometry since the bubble surface is very close to the wall in the second part of the cycle. Starting from given initial conditions, the subsequent bubble and motor displacements depend only on the bubble radius (see figures on the right of each panel of Figure~\ref{fig:parametricChemicals}), and hydrodynamics and geometry are observed to fully determine the total displacement of the motor, $\Delta z_c$.

In contrast, the chemical problem, set by the properties $H^{cc}$ and $\beta$, does affect the bubble growth rate by setting the amplitude of the dissolved gas flux, $Q$, and the duration of the bubble cycle, $\Delta T$, is therefore set by the chemical and diffusion dynamics. This decoupling between the hydrodynamic and chemical problems obviously breaks down when the bubble is unable to exit the cone (e.g.~for large values of $H^{cc}$).

\subsection{Microrocket displacement and average velocity for different opening angles}
\begin{figure}[h]
\center
\includegraphics[width=\textwidth]{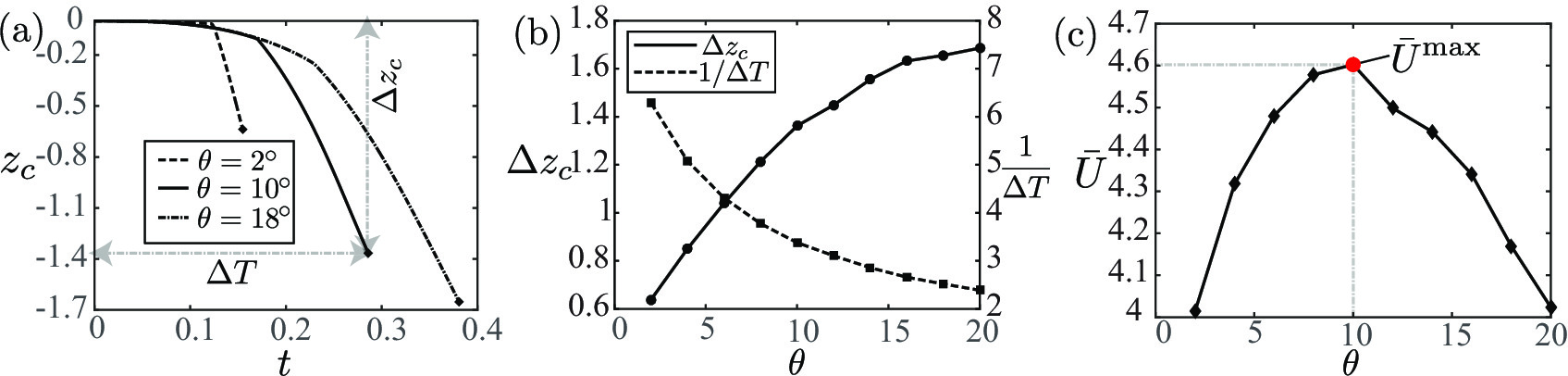}
\caption{Impact of the value of the opening angle, $\theta$, on the propulsion over one cycle for $H^{cc}=0.1$ and $\beta=1$. (a) Evolution with time of the microrocket displacement for different opening angles. (b) Influence of the opening angle, $\theta$, on the total cone displacement over the cycle, $\Delta z_c$, and bubble ejection frequency, $1/\Delta T$. (c) Corresponding evolution of the average cone velocity, $\bar U=\Delta z_c/\Delta T$. The maximum value is reached for $\theta=10^\circ$.}
\label{fig:coneAngleAdim}
\end{figure}

With this understanding, we now turn to the role of motor geometry and study the dependence of the average velocity, $\bar{U}=\Delta z_c/\Delta T$, on the opening angle, fixing the values $H^{cc}=0.1$ and $\beta=1$. The cone displacement is plotted as a function time for different opening angles $\theta$ on Figure~\ref{fig:coneAngleAdim}. The total displacement achieved over one bubble cycle is observed to increase with the opening angle, $\theta$, whereas the ejection frequency defined as $1/\Delta T$ decreases with $\theta$ (Figure~\ref{fig:coneAngleAdim}b). As a result, the average velocity $\bar{U}$ becomes negligible for small and large opening angles, because either the displacement is too small (small angles) or the cycle period diverges (large angles). As a result, an optimal opening angle $\theta_\textrm{opt}=10^\circ$ is identified for which $\bar U$ is maximum, as shown in Figure~\ref{fig:coneAngleAdim}c.

\subsection{Optimal microrocket design}
\label{sec:optimal}

\begin{figure}[h]
\center
\includegraphics[width=\textwidth]{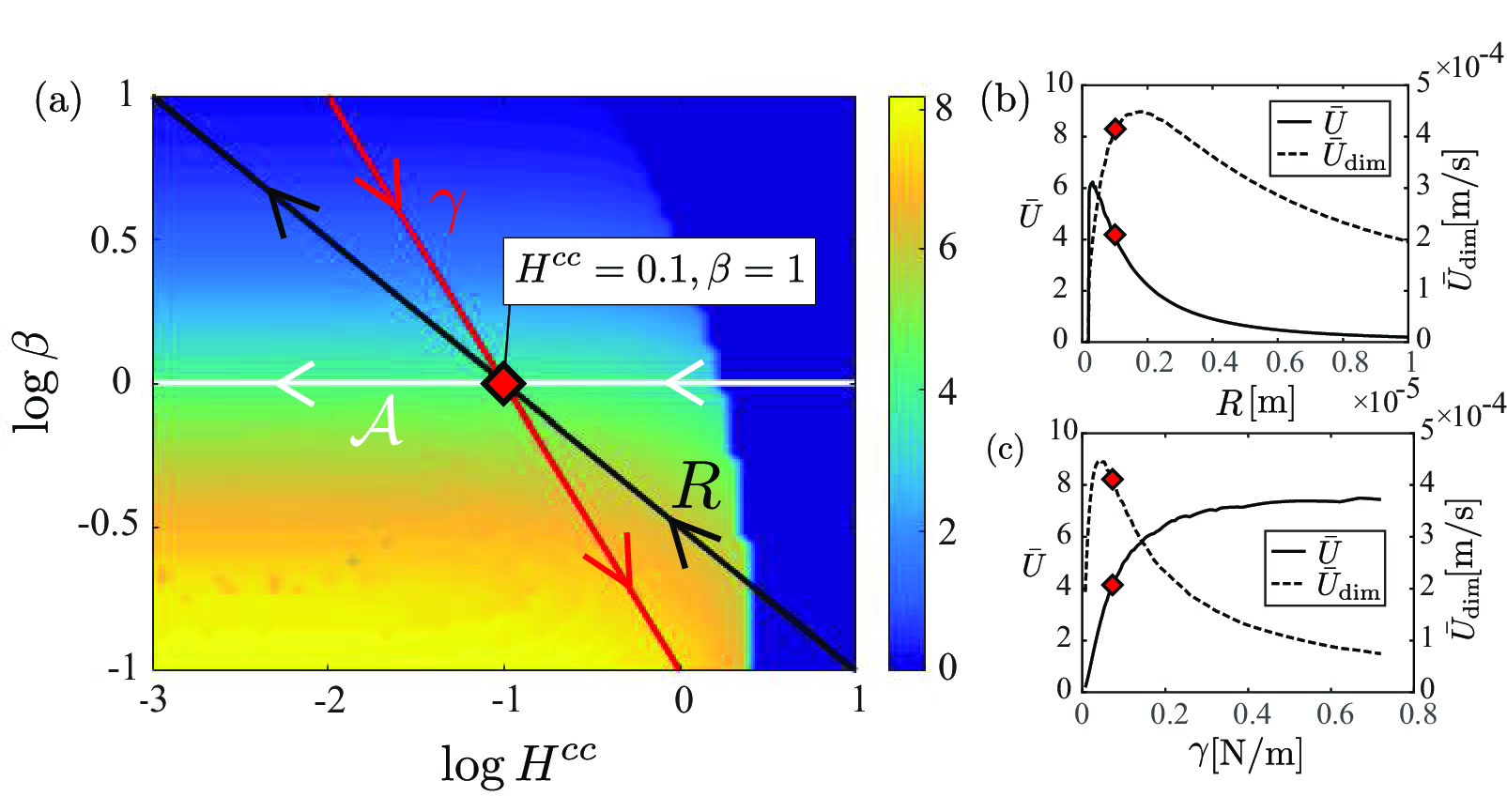}
\caption{Effect of the microrocket radius, $R$, chemical activity, $\A$, and surface tension, $\gamma$, on the propulsion: (a) Average microrocket velocity (non-dimensional) versus $H^{cc}$ and $\beta$; the red diamond corresponds to the reference conditions $H^{cc}=0.1$ and $\beta=1$, and coloured lines correspond to variations of the indicated dimensional parameter, all other remaining fixed. (b,c) Evolution of the average microrocket velocity, non-dimensional, $\bar{U}$, and dimensional, $\bar{U}_\text{dim}$, with (b) motor radius $R$ and (c) surface tension, $\gamma$, around the reference conditions denoted by a red diamond (see panel a).}
\label{fig:parametricChemicalsColorMap}
\end{figure}

In Figure~\ref{fig:parametricChemicalsColorMap}a we extend the results of Section~\ref{sec:results_chem} and plot the variation of the average velocity with the physico-chemical parameters $H^{cc}$ and $\beta$. Given the results of Figure~\ref{fig:parametricChemicals}, it comes as no surprise that the average velocity is a decreasing function of both $H^{cc}$ and $\beta$, since an increase in either of those parameters effectively increases the bubble cycle period until it becomes infinite (i.e.~the bubble stops inflating before reaching the cone exit).

Although this map provides useful information and physical insight, since both $H^{cc}$ and $\beta$ combine different parameters that are critical in experimental applications, it is not sufficient to disentangle the role of dimensional characteristics such as the cone radius, $R$, the chemical activity, $\A$, or surface tension, $\gamma$, which can be tuned during the fabrication process of the motor ($R$, $\A$) or by using surfactants ($\gamma$). In Figure~\ref{fig:parametricChemicalsColorMap}a, we overlap lines following the variations $R$, $\gamma$ and $\A$, respectively, all other parameters being held constants and equal to those introduced in section~\ref{sec:models}, starting from $H^{cc}=0.1$ and $\beta=1$ for which a motor radius $R=1\mu\m$ and surface tension $\gamma=7.2\times 10^{-2}\,N\,\,\m^{-1}$ lead to an average dimensional velocity $\bar{U}_\textrm{dim}=4.2\times 10^{-4} \m\,\,\s^{-1}$ (we refer to this in the following as the ``reference conditions''). The variations of the corresponding average velocity, $\bar{U}$, along these lines are shown in Figure~\ref{fig:parametricChemicalsColorMap}b-c, together with the corresponding dimensional velocity, $\bar{U}_\text{dim}$. One should note that the motor radii maximizing the non-dimensional and dimensional average velocities differ slightly, the former reaching its peak for a radius slightly larger than the reference configuration, while the latter peaks at lower values of $R$. Similarly, the non-dimensional velocity increases monotonically with surface tension $\gamma$, while the dimensional velocity presents a maximum at intermediate values of $\gamma$. This is a result of the reference velocity scale chosen here, namely $3R\A \R T_0/4\pi\gamma$ so that $\bar U_\textrm{dim}\sim (R/\gamma)\bar U$. Similarly, since $\bar{U}$ remains almost constant with $\A$ (see figure~\ref{fig:parametricChemicalsColorMap}a), $\bar{U}_\text{dim} \sim \A \bar{U}$ increases monotonically with 
$\A$  as is observed  experimentally~\cite{li2011dynamics}.

Focusing on the influence of the microrocket size, an increase of $R$ leads to a decrease in the non-dimensional velocity $\bar{U}$ as it inhibits bubble growth ($\beta$ is increased, corresponding to an increase in the relative influence of the ambient pressure on the bubble inner pressure and surface concentration). But increasing $R$ also leads to a larger dimensional velocity for fixed $H^{cc}$ and $\beta$. The competition of these two mechanisms results to negligible values of the dimensional velocity for both small and large $R$, and in the existence of an optimal motor radius.

\begin{figure}[h]
\center
\includegraphics[width=\textwidth]{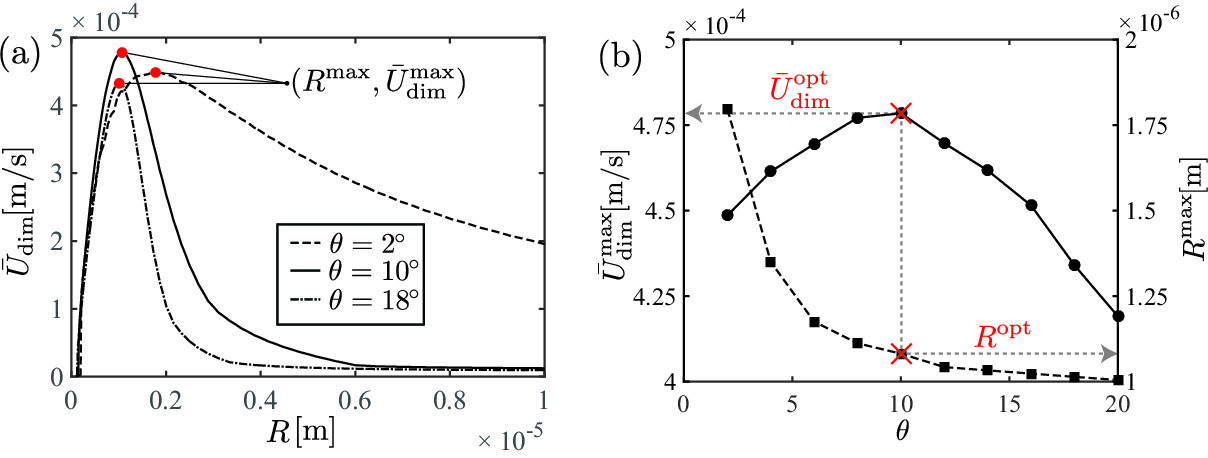}
\caption{Optimal microrocket design: (a) Evolution of the average dimensional velocity with the cone radius around the reference conditions for different opening angles. (b) Evolution with cone opening angle $\theta$ of the average velocity, $\bar{U}^\text{max}_\text{dim}$, and corresponding cone radius at which the maximum is attained, $R^\text{max}$.}
\label{fig:coneAngle}
\end{figure}

We now repeat such a numerical experiment for different opening angles $\theta$, retrieving a family of curves $(R,\bar{U}_\text{dim})$ (Figure~\ref{fig:coneAngle}a) and identifying for each value of $\theta$, the optimal dimensional velocity, $\bar U_\textrm{dim}^\textrm{max}$, and the optimal radius, $R^\textrm{max}$. As the cone angle increases, the maximum average velocity $\bar{U}^\text{max}_\text{dim}$ is reached for a smaller cone radius, i.e.~$R^\textrm{max}$ is a decreasing function of $\theta$ (Figure~\ref{fig:coneAngle}b). In order to work in optimal conditions, larger microrockets are thus required for smaller opening angles. Furthermore, a global optimum $\bar{U}_{\text{dim}}^{\text{max}}=4.78 \times 10^{-4} $m~s$^{-1}$ is identified for $\theta=10^\circ$, to which corresponds a cone radius $R^\textrm{opt}\approx 1.2 \times 10^{-6} \m$.

\subsection{Perspectives: many bubbles interaction}
\label{sec:perspectives}

In this section, we show preliminary results for the propulsion of the microrocket  in the case where many bubbles are present in the cone. This scenario is of interest for real applications where a train of closely spaced bubbles is observed to exit the microrocket. 

\begin{figure}[t]
\center
\includegraphics[width=\textwidth]{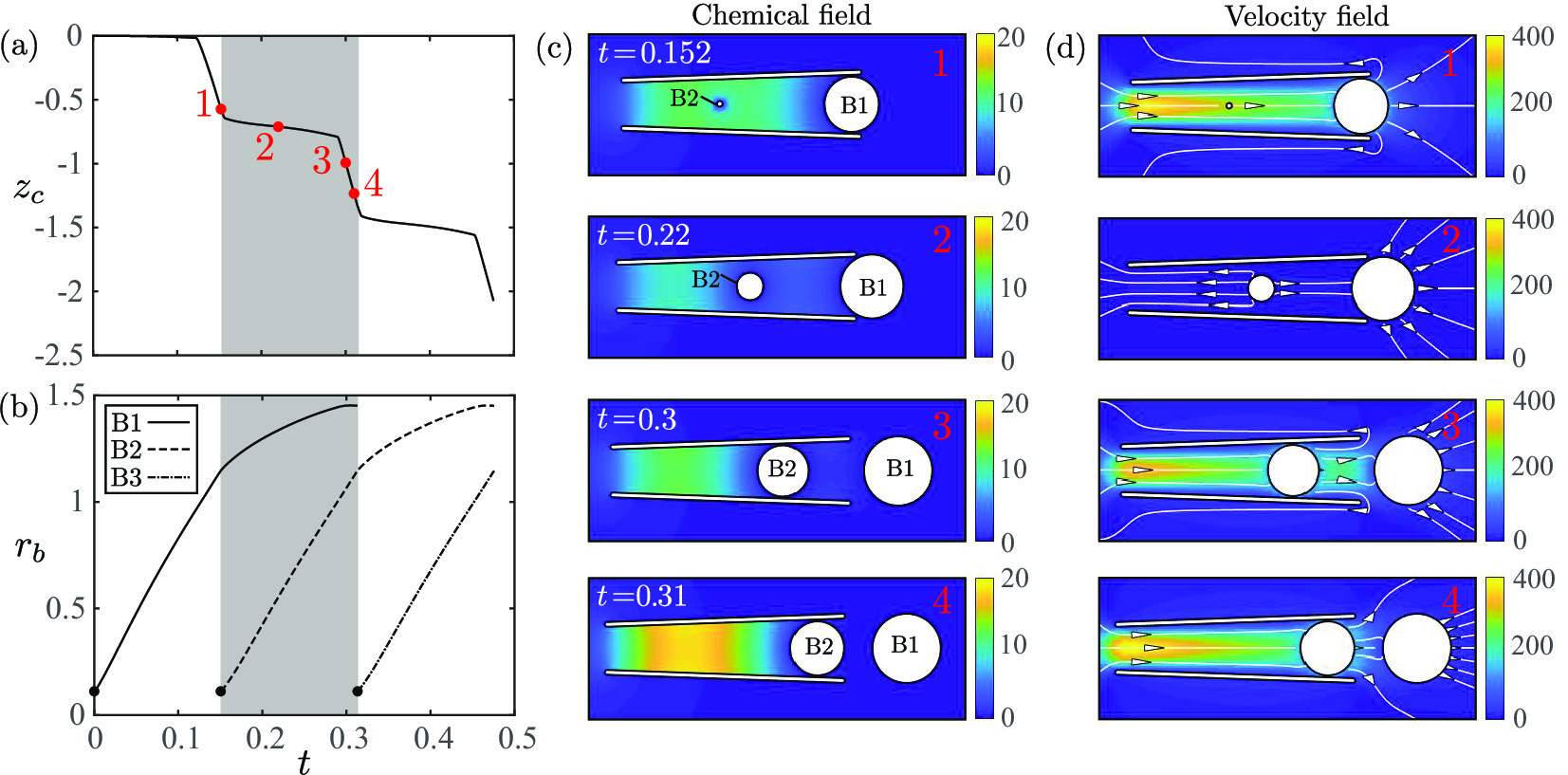}
\caption{Chemical concentration and bubble dynamics when many bubble are present in the cone for the parameters $H^{cc}=0.1$, $\beta=1$ and $\theta=2^\circ$: (a) Time-dependence of the cone position, with the shaded region highlighting one bubble cycle; (b) Time-dependence of the bubble radius, $r_b$, with each line describing a different bubble; (c,d) Snapshots of the dissolved gas concentration and velocity field (streamlines and intensity) at the four instants shown in panels (a)}
\label{fig:twoBubbles}
\end{figure}

The results presented below use  the same parameters as  in section~\ref{sec:bubble_cycle} and more systematic investigations will be addressed in our future work. In  order for the problem to remain tractable, we  also enforce two additional rules:

\begin{itemize}

\item After the nucleation of the first bubble, which occurs as explained  in section~\ref{sec:bubble_nucleation}, subsequent bubbles nucleate on the axis when the concentration exceeds a threshold value, $c_N$, with radius $r_b(0) = 2H^{cc}/(c_N -\beta H^{cc})+0.1$, similarly to section~\ref{sec:bubble_nucleation}. The value $c_N=20$ was chosen as an illustration;  note that this value does not derive from thermodynamic considerations, as it is challenging to precisely determine how bubble nucleation occurs on catalyst surfaces~\cite{lv2017growth}. Instead, the value of $c_N$ is a tuning parameter  selected in order to obtain a bubble ejection frequency compatible with the one observed in experiments~\cite{manjare2013bubble}.

\item The presence of multiple bubbles increases significantly the computational complexity of the system, thus the number of coexisting bubble is limited to two. Namely, when a third bubble nucleates, we eliminate the bubble farthest away from the cone. We expect that this assumption will not strongly affect these preliminary results, because eliminated bubbles have already exited the cone and therefore weakly contribute to the cone displacement and the chemical environment within the cone. In fact, when out of the cone, bubbles receive a small chemical flux (or even shrink, see Figure~\ref{fig:twoBubbles}b) which leads to a small fluid displacement (see intensity of the fluid velocity in Figure~\ref{fig:twoBubbles}d, snapshots $3$ and $4$).

\end{itemize}
Under these two rules,  we illustrate our computational results for $\theta=2^\circ$, $H^{cc}=0.1$ and $\beta = 1$ in Figure~\ref{fig:twoBubbles}. The first part of the simulation is identical to the results illustrated in section~\ref{sec:bubble_cycle}, namely a slow cone displacement when the first bubble B$1$ is not confined followed by a sharp cone acceleration during the confined phase. At time $t=0.152$ a second bubble B$2$ nucleates because max$(c(z,0))>c_N$. 
Running very long simulations, we  observe that from this instant the dynamics repeats periodically roughly every $0.16$ unit of time (see video $\#2$ in the Supporting Information). In order to investigate the microrocket dynamics, we can therefore focus on the periodic dynamics that defines the bubble cycle (shaded region in Figure~\ref{fig:twoBubbles}a-b and corresponding snapshots in Figure~\ref{fig:twoBubbles}c-d).

The nucleation of bubble B$2$ lowers the local concentration but does not strongly affect the flow because of its small size compared to B$1$. In fact, B$1$ keeps translating toward the larger opening, generating a flow that drags B$2$ along. Subsequently, B$1$ exits the cone and B$2$ inflates but it does not yet feel the confinement: in this phase the cone displaces slowly because none of the two bubbles are geometrically confined. When B$2$ becomes larger, it translates because of the geometrical confinement and  pushes B$1$ out of the cone. Because B$1$ is now completely outside the cone, it absorbs less chemical flux and eventually shrinks. Finally, B$2$ keeps inflating and translating while the concentration in the left part of the cone recovers its original higher level. Shortly after $t=0.31$, a third bubble B$3$ nucleates and the bubble cycle starts again. The computed average microrocket velocity is $\bar{U}\approx 4.6$, which is larger than that obtained in section~\ref{sec:bubble_cycle}, where $\bar{U}\approx 4$. This difference is mostly due to the fact that, when B$2$ is inside the cone but is not geometrically confined (see snapshot $2$ in Figure~\ref{fig:twoBubbles}c-d), B$1$ is still partially confined and provides thrust to the microrocket. This situation is considerably different from the one-bubble case, where almost no thrust is provided while the bubble is not confined. In fact, the displacement attained when the bubbles are not strongly confined is considerably smaller in the one-bubble case ($\Delta z_c=0.017$ from $t=0$ to $t\approx 0.12$) compared to the two-bubble case ($\Delta z_c=0.163$ from $t=0.155$ to $t\approx 0.29$).

\section{Conclusion}
In summary, in this paper we have used numerical simulations to develop a joint chemical and hydrodynamic analysis of the bubble growth within a conical catalytic microrocket and of the associated bubble and microrocket motion. Our computations have revealed number of important physical features. First, we have found that most of the displacement of the microrocket is attained when the bubble is strongly confined by the conical-shaped swimmer. Second, we have shown that the chemical and the hydrodynamic problem can be decoupled: the chemical problem determines the bubble ejection frequency while the hydrodynamic problem determines the microrocket displacement. Finally, we have systematically studied the microrocket swimming velocity, finding the optimal cone shape and size which maximize it.

In our future work, we plan to explore the relevance of the bubble deformation and the interaction between many bubbles, both  relevant to real applications where a lot of bubbles are seen to be emitted close to each other. Preliminary computational results seem to indicate that, although the presence of many bubbles slighly modify the quantitative performance of the motor, it does not alter the qualitative picture obtained for the one-bubble case where a periodic bubble cycle establishes and essentially all the microrocket displacement is obtained when bubbles are geometrically confined by the conical-shaped swimmer. Overall, our results shed light on the fundamental chemical and hydrodynamic processes of the propulsion of catalytic conical swimmers and will allow the experimental design of optimal bubble-propelled microrockets.


\section*{Acknowledgements}

This project has received funding from the Swiss National Science Foundation (SNFS) with the Doc.Mobility Fellowship P1ELP2$\_$172277 (G.G.) and the European Research Council (ERC) under the
European Union's Horizon 2020 research and innovation program under grant agreements 714027 (S.M.), 682754 (E.L.) and 280117 (F.G.). The computer time has been provided by SCITAS at EPFL.

\section*{Keywords}
microswimmer, microrocket, self-propulsion, catalytic swimmer, numerical simulations

\bibliographystyle{unsrt}

\end{document}